\documentclass[a4paper,twocolumn,fleqn]{article}

\usepackage{txfonts}               
\usepackage{pfr}                   
\usepackage{graphicx}
\usepackage{color}

\begin{document}

\title{Physics-Informed Machine Learning Approach to Modeling Line Emission from Helium-Containing Plasmas}


\author{Shin Kajita}

\affiliation{Graduate School of Frontier Sciences
The University of Tokyo
5-1-5 Kashiwanoha, Kashiwa, Chiba 277-8561, JAPAN
}
\date{}

\email{kajita@k.u-tokyo.ac.jp}

\begin{abstract}
The helium I line intensity ratio (LIR) method is used to measure the electron density ($n_e$) and temperature ($T_e$) of fusion-relevant plasmas. Although the collisional-radiative model (CRM) has been used to predict $n_e$ and $T_e$, recent studies have shown that machine learning approaches can provide better measurements if a sufficient dataset for training is available.
This study investigates a hybrid neural network approach that combines CRM- and experiment-based models. Although the CRM-based model alone exhibited negative transfer in most cases, the ensemble model modestly improved the prediction accuracy of $T_e$.
Notably, in data-limited scenarios, the CRM-based model outperformed the others for $T_e$ prediction, highlighting its potential for applications with constrained diagnostic access.
\end{abstract}


\DOI{}

\maketitle  

The helium (He) I line intensity ratio (LIR) method has been used to measure the electron density, $n_e$, and temperature, $T_e$, in He-containing plasmas~\cite{Schweer1992JNM,Agostini2020RSI,delaCal2001PPCF,Ohno2010CPP,Kimata2022AIP}.
Because He atoms are produced by the deuterium-tritium fusion reaction, this method could be an important diagnostic tool in future fusion devices, where plasma diagnostics will be much more limited than in current experimental fusion reactors.
The LIR method compares experimental data with collisional radiative model (CRM) calculations, which demonstrate the dependence of the He~I line emission on $n_e$ and $T_e$~\cite{Goto2003JQSRT}.

However, explaining all the processes that determine the population distribution is not so simple. Linear plasma devices have reported that radiation transport of resonance lines in the ultra-violet (UV) range can significantly disturb the population distribution, particularly the $^1$P states~\cite{Sasaki1996RSI,Kajita2006POP,Kajita2009POP,Kajita2009POP2,Iida2010RSI,Sawada2010PFR}. 
This effect depends not only on the neutral pressure, but also on the neutral temperature~\cite{Nishijima2007PPCF} and the spatial emission profile~\cite{Kajita2011RSI,Iida2010POP}.
In low-temperature recombining plasmas, some discrepancies have been identified, and the effect of metastable atoms, which were produced mainly by the recombination processes, can influence the population distribution~\cite{Kajita2018POP}.
In high-density plasmas in Magnum-PSI, the reason for the discrepancy between the CRM and experiments cannot be explained under certain conditions~\cite{Kajita2020AIPAdv}.

Recent works in linear plasma devices have revealed that machine learning approaches to learn the relation between the optical emission spectroscopy (OES) data and $n_e$ and $T_e$ work well on the measurement of $n_e$ and $T_e$~\cite{Nishijima2021RSI,Kajita2021PPCF,Kajita2022NME,Kajita2023FED}. 
Unfortunately, however, this method does not take advantage of CRM knowledge. 
The question is whether it is possible for the machine learning approach to use CRM data to reduce the prediction error.
In this work, we adopt a physics-informed machine learning (PIML) approach~\cite{Karniadakis2021NRP}, where the simulation results from a CRM are incorporated into a neural network (NN) via pre-training.
We demonstrate cases where the PIML with CRM data can reduce prediction error.

This study uses the same dataset previously used~\cite{Kajita2024PPCF} from the linear plasma device Magnum-PSI~\cite{VANECK2014FED}, which can produce high-density ($n_e<10^{21}$~m$^{-3}$) and low-temperature ($T_e<5$~eV) plasmas in a steady state.
The dataset comprises of the optical emission spectroscopy (OES) data from He and He-hydrogen (H) mixed plasmas, as well as laser Thomson scattering (TS) data of $n_e$ and $T_e$. The spectroscopy output CCD data were preprocessed using Method~1 in Ref.~\cite{Kajita2024PPCF}, in which the CCD data around 16 line-emissions observed in pure He plasmas and the edge data in the wavelength direction are used as inputs. In total, there are 1826 data points comprising 49 different discharge conditions and approximately 40 spatial points in a radial profile. 

\begin{figure}[tbp]
\centering
\includegraphics[width=7cm]{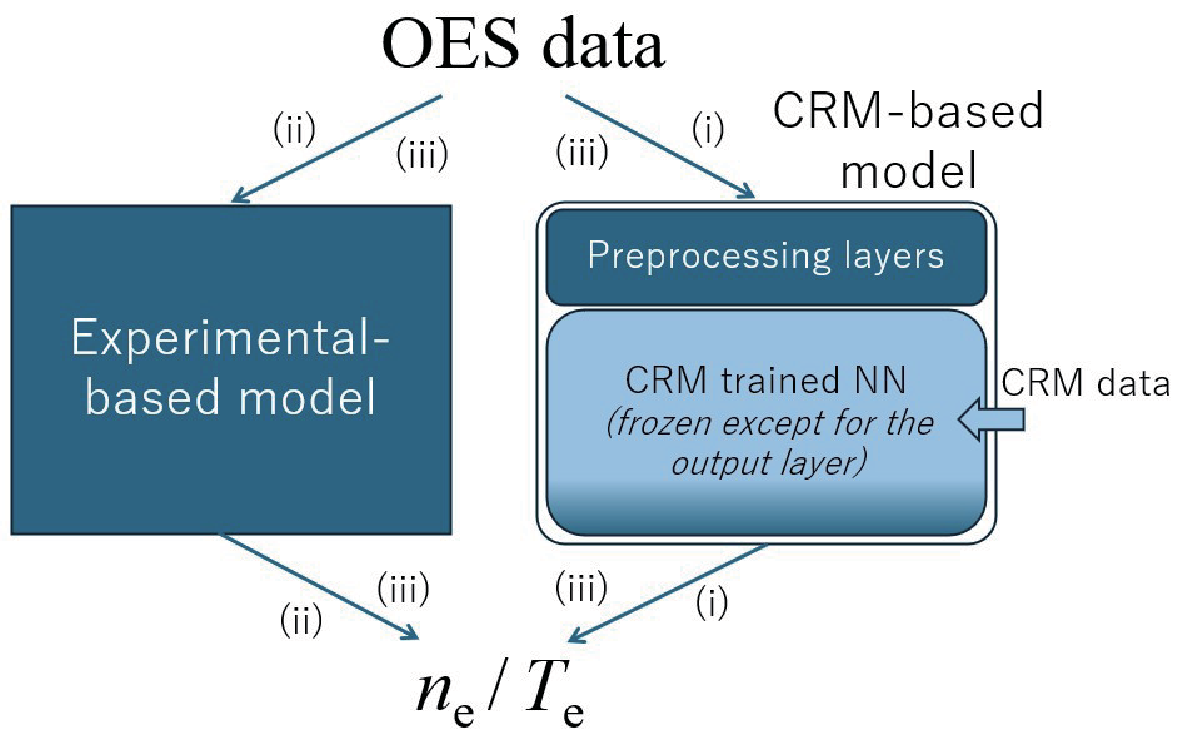}
\caption{A schematic of the models used.}
\label{Fig:Model}
\end{figure}

Figure~\ref{Fig:Model} shows a schematic of the NN models used in this work. There are three models, i.e., (i) CRM-based model, (ii) exp-based model, and (iii) ensemble model. 
In the CRM-based model, first a CRM was used to compute the reduced population coefficients, and the relationship between the coefficients and $n_e$/$T_e$ is trained. 
In formulation II in the CRM, where quasi-steady-state approximation is used, the excited level population is described as~\cite{Goto2003JQSRT}
\begin{eqnarray}
n(p)=R_0(p) n_e n_i + R_1(p) n_e n_0, \label{eq}
\end{eqnarray}
where $R_0(p)$ and $R_1(p)$ are the reduced population coefficients, $n_i$ is the ion density, and $n_0$ is the ground state He atomic density. The first and second terms in Eq.~(\ref{eq}) correspond to the recombining and ionizing components, respectively. 
The ranges used for the CRM calculations are $1.0 \times 10^{18} \leq n_e \leq 2.0 \times 10^{21}$~m$^{-3}$, $0.1 \leq T_e \leq 5.1$~eV. In addition, the effect of radiation trapping was taken into account using the optical escape factor~\cite{Kajita2024JPD}, which depends on the optical depth and is a function of the neutral density and the radius of a cylinder, $R$. The calculation was performed in the range of $0 \leq n_0R \leq 2.6 \times 10^{20}$~m$^{-2}$. 
In this study, we performed a full grid scan using 52 points for $T_e$, 77 points for $n_e$, and 16 points for $n_0R$. As a result, a total of 52$\times$77$\times$16 = 64,064 combinations were calculated. The reduced population coefficients for the upper states of the line emissions at 728.1, 706.5, 501.6, 388.9, 667.8, 492.2, 447.1, 438.8, and 402.6~nm, i.e., 3$^1$S, 3$^3$S, 3$^1$P, 3$^3$P. 3$^1$D, 4$^1$D, 4$^3$D, 5$^1$D, and 5$^3$D, are obtained and used for the CRM-based model.
From Eq.~(\ref{eq}), the ideal inputs to the NN are $R_0(p)\, n_e\, n_i$ and $R_1(p)\, n_e\, n_0$. However, for practical implementation, $R_0(p)\, n_e^2$ was used to represent the first term by assuming $n_i = n_e$, and $R_1(p)\, n_e$ was used to represent the second term, omitting the unknown $n_0$. The NN consisted of five hidden layers with 256, 256, 128, 64, and 16 neurons, respectively.

To construct the CRM-based model, we first developed a NN trained solely on CRM computation data, using 80\% for training and the remaining 20\% for validation. The resulting models achieved mean percentage errors (MPEs) of 0.18\% for $n_e$ and 0.25\% for $T_e$, both of which are substantially lower  than typical measurement errors. 
To integrate this with the experimental OES data, we introduced preprocessing layers including a hidden layer with 64 nodes. During training with experimental data, all layers of the CRM-trained NN were frozen except for the final output layer.

For (ii) exp-based model, a NN with five hidden layers with 256, 256, 128, 64, and 16 neurons, respectively, was used. 
For (iii) ensemble model, the trained CRM-based NN and exp-based NN are combined with an additional dense layer of 16 nodes. Except for the added last layer, the NN was frozen, and the trained model was used.

The OES data, which was preprocessed using the Method~1 in Ref.~\cite{Kajita2024PPCF}, the reduced coefficients, and $n_e$/$T_e$ are converted to logarithmic scale, and a feature scaling was performed on all the data before being used for training.
We use Tensorflow~\cite{abadi2016tensorflow} and Keras~\cite{chollet2015keras} libraries to implement these models.

\begin{figure}[tbp]
\centering
\includegraphics[width=7cm]{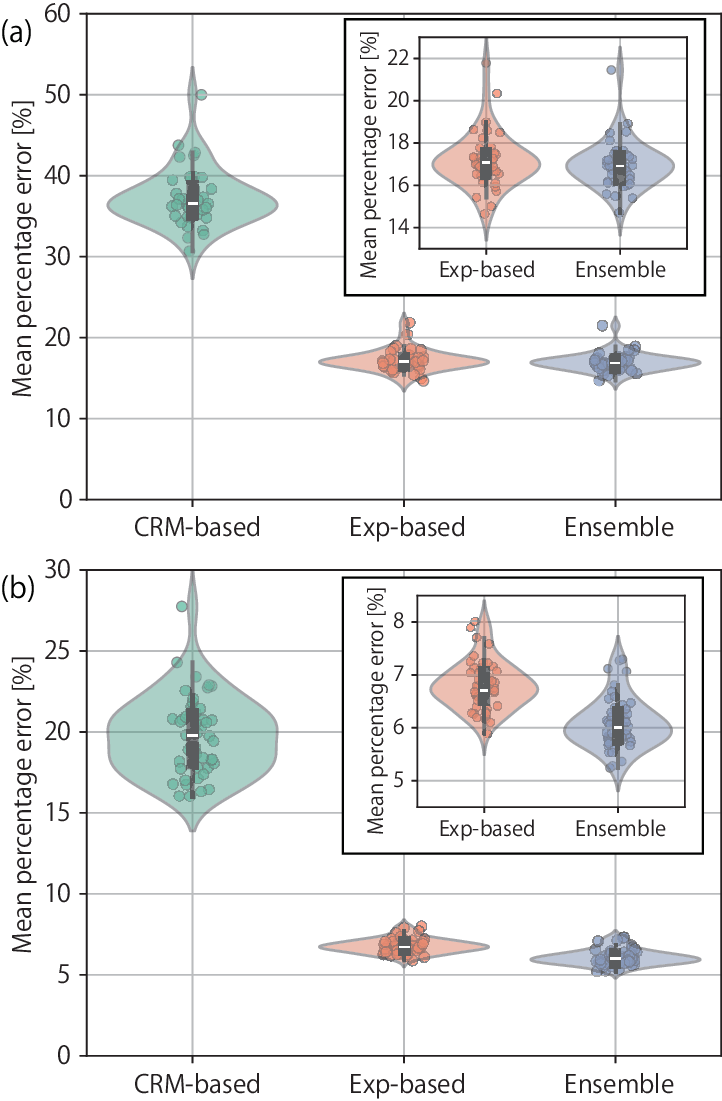}
\caption{Comparisons between the three models using MPEs for the prediction of (a) $n_e$ and (b) $T_e$. The data were randomly split into training and test sets in an 80:20 ratio from a total of 1826 data points.}
\label{Fig:Method2}
\end{figure}

Figure~\ref{Fig:Method2}(a,b) shows comparisons between the three models using MPEs for $n_e$ and $T_e$, respectively. 
The training data was randomly selected from all data points, and 80\% was used for training and the remaining 20\% was used for testing. 
The calculation was repeated 50 times, selecting new data each time; each point in Fig.~\ref{Fig:Method2} corresponds to an MPE for an optimized NN, and the same set of data is used for the three models. 
The CRM-based model has larger errors than the exp-based or ensemble models for both $n_e$ and $T_e$. 
The CRM-based model is a transfer learning model, and the results show that negative transfer~\cite{Zhang2023IEEE}, i.e., interference of the previous knowledge (CRM data) with the new learning and undesirably reduces the learning performance in the target domain, occurred. 
Although the reason for the negative transfer is not clear, it could be caused by the fact that the CRM calculation does not agree with the experimental data, as discussed earlier~\cite{Kajita2020AIPAdv}, probably due to line integration effects, molecular effects, transport of metastable states, etc.
When comparing the exp-based model and the ensemble model, the difference between them was not statistically significant at the 0.05 level for $n_e$ ($t = 1.85$, $p = 0.072$). 
That is, the knowledge of CRM did not contribute to the performance of $n_e$ prediction.
In contrast, for $T_e$, although the difference in the MPE is less than 1\%, the difference was statistically significant ($t = 10.30$ and $p = 7.38 \times 10^{-14}$), indicating a strong discrepancy between the experimental and ensemble values. 
Even if the CRM model is individually worse, it may still capture aspects of the $T_e$ prediction that the experimental model does not. The ensemble can learn to down-weight unhelpful CRM inputs and exploit useful patterns.

\begin{figure}[tbp]
\centering
\includegraphics[width=7cm]{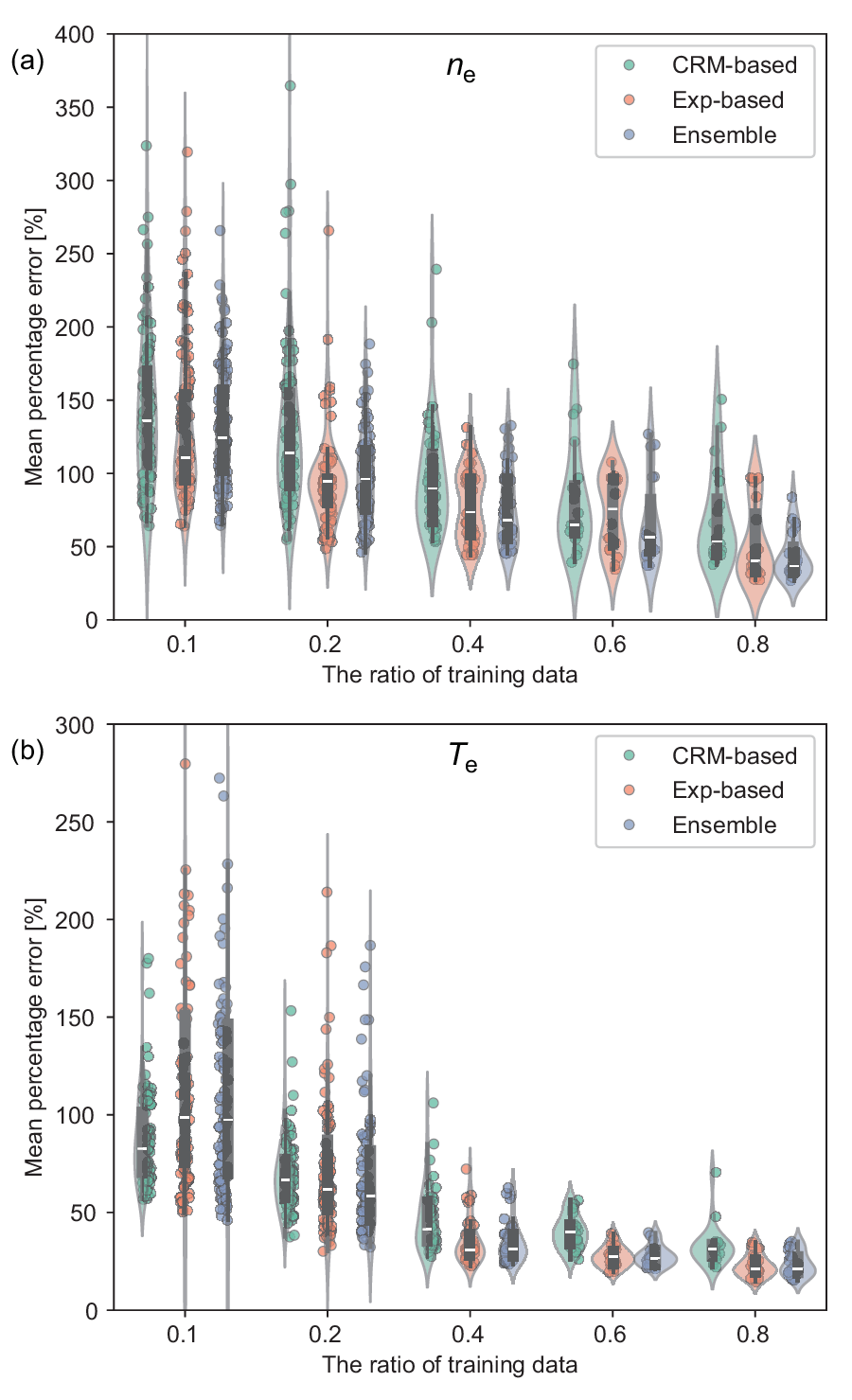}
\caption{The prediction error in (a) $n_e$ and (b) $T_e$ for the three models with different train-test split ratios between 80:20 and 10:90. The data were split into training and test sets by discharge unit from a total of 49 discharges.}
\label{Fig:Method1}
\end{figure}

One of the challenging issues is whether we could reduce the necessary data for training the model by using CRM data. To see the impact, here, we separate the training and validation data in the unit of discharge to consider realistic situation~\cite{Kajita2022NME}. 
The data from 49 discharges were randomly split by discharge unit into training and test sets, with the train-test ratio varied from 80:20 to 10:90. The average prediction errors in $n_e$ and $T_e$ for the three models are shown in Fig.~\ref{Fig:Method1}(a,b), respectively. For instance, at a 10:90 train-test split, five discharges were used for training, and the remaining 44 discharges were used for validation. This random selection process was repeated over 20 times for each split ratio, and the distribution of the average prediction errors is analyzed.
Unlike Fig.~\ref{Fig:Method2}, the errors are larger, because the unit of separation in Fig.~\ref{Fig:Method1} is discharge, i.e. training data does not contain the same discharge as the test data. 
The average MPEs of the CRM- and exp-based models at a train-test ratio of 80:20 were 52.3\% and 69.0\%, respectively, for $n_e$, and 32.9\% and 22.8\%, respectively, for $T_e$. These values are much greater than those in Fig.~\ref{Fig:Method2} (37.3\% and 17.2\%, respectively, for $n_e$, and 19.7\% and 6.8\%, respectively, for $T_e$).
Like Fig.~\ref{Fig:Method2}, the exp-based model and the ensemble model have better performance than the CRM-based model in most cases, and the difference between the exp-based model and the ensemble models is marginal. 
For $n_e$ (Fig.~\ref{Fig:Method1}(a)), the average MPE is $\sim$100\% or higher when the training data ratio is 20\% or less. The CRM-based model has a higher average MPE than the other two models, suggesting that negative transfer also occurred for all $n_e$ cases.
However, for $T_e$, when the train-test split ratio is 10:90, where only five discharge data were used for training, while exp-based model and ensemble model did not work properly with the average MPE of $\sim$100\%, but the CRM-based model has lower average MPE of 87\%. 
The difference between the CRM-based model and exp-based model at the split ratio of 10:90 was statistically significant ($t=3.21$ and $p=0.0018$).
Therefore, it can be said that the CRM-based model has the advantage of having a better quality in extreme cases where the available amount of data is significantly limited.

In this study, we developed and compared three NN models for predicting $n_e$ and $T_e$ from high-density and low-temperature He containing plasmas: a CRM-based model using transfer learning, a model trained directly on experimental data (exp-based model), and an ensemble model combining both. 
While the CRM-based model alone exhibited negative transfer and underperformed relative to the exp-based model, the ensemble model achieved slightly improved accuracy in predicting $T_e$, with statistical significance. No significant improvement was observed for $n_e$. Notably, under extreme data-scarce conditions, the CRM-based model demonstrated better performance than models trained solely on experimental data for $T_e$ measurement, highlighting its potential value in future applications where limited diagnostic access is expected. 
In the future, it is expected that the CRM-based model will be improved by introducing the formulation I of the CRM, where the transport of He atoms in the metastable state can be taken into account, which can complement data-driven approaches and increase diagnostic robustness, especially in low-data regimes.

The author thanks Dr. K. Fujii from Oak Ridge National Laboratory for fruitful discussion. The data used in this study were obtained at the DIFFER Institute, and the author would like to express their sincere gratitude for the support and access provided.
This work was supported by JSPS, Japan KAKENHI (24H00201 and 25H00615), NIFS Collaboration Research program, Japan (NIFS25KFGT001).


\end{document}